\documentclass[reprint,superscriptaddress,showpacs,amsmath,amssymb,aps,prl,floatfix,lengthcheck]{revtex4-1}
\usepackage{graphicx}% Include figure files
\usepackage{dcolumn}% Align table columns on decimal point
\usepackage{bm}% bold math
\usepackage{hyperref}% add hypertext capabilities

\begin{document}
%\linenumbers
%\preprint{APS/123-QED}

\title{Effect of a magnetic field on the quasiparticle recombination in superconductors}
\author{Xiaoxiang Xi}
\affiliation{Department of Physics, University of Florida, Gainesville, Florida 32611, USA}
\affiliation{Photon Sciences, Brookhaven National Laboratory, Upton, New York 11973, USA}
\author{J. Hwang}
\affiliation{Department of Physics, University of Florida, Gainesville, Florida 32611, USA}
\affiliation{Department of Physics, Sungkyunkwan University, Suwon, Gyeonggi-do 440-746, Republic of Korea}
\author{C. Martin}
\affiliation{Department of Physics, University of Florida, Gainesville, Florida 32611, USA}
\author{D. H. Reitze}
\affiliation{Department of Physics, University of Florida, Gainesville, Florida 32611, USA}
\author{C. J. Stanton}
\affiliation{Department of Physics, University of Florida, Gainesville, Florida 32611, USA}
\author{D. B. Tanner}
\affiliation{Department of Physics, University of Florida, Gainesville, Florida 32611, USA}
\author{G. L. Carr}
\affiliation{Photon Sciences, Brookhaven National Laboratory, Upton, New York 11973, USA}
\date{\today}

\begin{abstract}
Quasiparticle recombination in a superconductor with an \textit{s}-wave gap is typically dominated by a phonon bottleneck effect. We have studied how a magnetic field changes this recombination process in metallic thin-film superconductors, finding that the quasiparticle recombination process is significantly slowed as the field increases. While we observe this for all field orientations, we focus here on the results for a field applied parallel to the thin film surface, minimizing the influence of vortices. The magnetic field disrupts the time-reversal symmetry of the pairs, giving them a finite lifetime and decreasing the energy gap. The field could also polarize the quasiparticle spins, producing different populations of spin-up and spin-down quasiparticles. Both processes favor slower recombination; in our materials we conclude that strong spin-orbit scattering reduces the spin polarization, leaving the field-induced gap reduction as the dominant effect and accounting quantitatively for the observed recombination rate reduction.
\end{abstract}
\maketitle

An excitation from the superconducting condensate requires finite energy (the energy gap 2$\Delta$) and produces two quasiparticles.  A quasiparticle excited to very high energy (compared to $\Delta$) quickly decays via a number of fast scattering processes to near the gap edge, where it recombines with a partner to form a Cooper pair. The pair's binding energy is emitted mainly as 2$\Delta$ phonons \cite{Ginsberg1962,Schrieffer1962,Tewordt1962}. The recombination process is delayed by a phonon bottleneck: each recombination-generated phonon can break another Cooper pair, causing energy to be trapped in a coupled system of 2$\Delta$ phonons and excess gap-edge quasiparticles \cite{Rothwarf1967,Schuller1975,Chi1981}. Quasiparticle recombination has been widely studied in both metallic superconductors, to investigate the non-equilibrium processes in the many-body BCS system \cite{Federici1992,Carr2000,Demsar2003,Beck2011}, and high-temperature superconductors, to gain new insight into the pairing mechanism \cite{Gedik2004,Kusar2008,Mertelj2009,Mihailovic1999}. Theories of the recombination process considered the reaction kinetics and interactions of quasiparticles and phonons \cite{Kaplan1976,Chang1977}; while experiments obtained the dependence of the quasiparticle lifetime on temperature, film thickness, and excitation strength \cite{Levine1968,Perrin1982}. A magnetic field is known to couple to the electron orbital motion and to align the spin; both effects weaken superconductivity \cite{Fulde2010}. The consequence of magnetic field on the quasiparticle recombination \cite{Holdik1985} has not been examined in detail by optical pump-probe methods. 

\begin{figure}[hb!]
\includegraphics[width=0.49\textwidth]{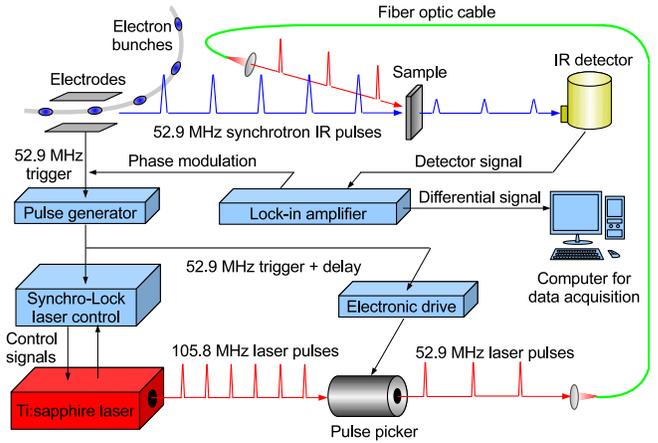}                
\caption{Experimental set-up. Electrons circulate in bunches in the synchrotron storage ring, generating pulses of far-infrared radiation with a repetition frequency of 52.9~MHz. The Ti:sapphire laser produces pulses with a repetition frequency of 105.8~MHz and a pulse picker selects every other pulse to match the synchrotron pulse pattern. The selected laser pulses are delivered over a fiber optic cable to the sample and the synchrotron pulse probes the photoinduced transmission at a fixed time delay afterward. To synchronize the synchrotron and laser pulses, the 52.9~MHz bunch timing signal from a pair of electrodes inside the synchrotron ring chamber is used by the Synchro-Lock laser control system as a reference for the laser pulse emission. The pulse generator introduces an adjustable delay between the laser and synchrotron pulses. The transmitted far-infrared light is detected by a bolometer detector and recorded on a computer.} 
\label{Fig1}
\end{figure}

We use a novel time-resolved laser-pump synchrotron-probe spectroscopic technique to study the quasiparticle recombination dynamics in superconducting thin films, under applied magnetic field. Samples studied include a 10~nm thick Nb$_{0.5}$Ti$_{0.5}$N film on a crystal quartz substrate and a 70~nm thick NbN film on a MgO substrate. These substrates are essentially transparent in the far-infrared spectral range. The films were grown by reactive magnetron sputtering, using NbTi cathode in Ar/N$_2$ gas for Nb$_{0.5}$Ti$_{0.5}$N and Nb cathode in N$_2$ gas for NbN. The two films have critical temperatures of 10.2 K and 12.8 K, and a zero-temperature, zero-field, energy gap 2$\Delta$ of 2.7~meV and 4.5~meV, respectively. Four-probe resistivity measurements with magnetic field parallel to the films determined their upper critical field to be greater than 20~T at $T\le3$~K. 

The samples were mounted in a $^4$He Oxford cryostat equipped with a 10~T superconducting magnet, and probed by far-infrared radiation produced in a bending magnet at beamline U4IR of the National Synchrotron Light Source, Brookhaven National Laboratory. The experiment, illustrated in FIG.~\ref{Fig1}, exploits the fact that the synchrotron radiation is emitted in $\sim$300~ps long pulses (governed by the electron bunch structure in the storage ring). We applied mode-locked near-infrared Ti:sapphire laser pulses ($\sim$2~ps in duration and $\sim$1.5~eV in photon energy) as the source for photoexcitation. The synchrotron probe beam measures the photoinduced optical properties due to the excess quasiparticles as a function of time delay relative to the arrival of the pump beam. The synchrotron pulse has a Gaussian profile with a FWHM of $\sim$300~ps, determining the time resolution of the experiment. At selected delay times $t$, we measure the spectrally integrated photoinduced transmission $S(t)\equiv -\Delta\mathcal{T}(t)$ over the spectral range spanning the superconductor’s energy gap ($\sim$3 meV). The spectral shape is determined primarily by the optical components carrying the beam, and the detector. 

If the laser were turned on and off to measure the photoinduced response, there would be a temperature modulation as well as the photoexcited quasiparticle modulation. To reduce these thermal effects we dither the laser pulse back and forth by a few tens of picoseconds at each delay setting, keeping the incident laser power constant. The dither is achieved by phase modulating the laser pulse using the internal oscillator of a lock-in amplifier. The directly obtained quantity is therefore a differential signal, $dS/dt$. This signal was detected using a B-doped Si bolometer in combination with the lock-in amplifier. Numerical integration yields the photoinduced transmission $S(t)$, which directly follows the excess quasiparticle density \cite{Lobo2005}. 

\begin{figure}[t]
\includegraphics[width=0.49\textwidth]{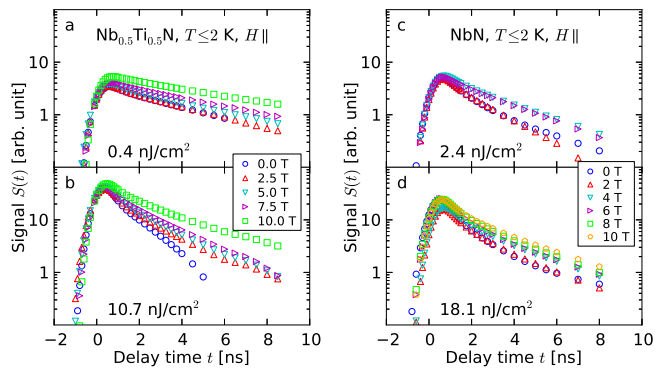}                
\caption{Photoinduced transmission $S(t)$ vs. time $t$ for Nb$_{0.5}$Ti$_{0.5}$N (a and b) and for NbN (c and d), all measured in parallel fields at $T\le2$~K. Low-fluence and high-fluence data are compared. Note the semilog scale; simple exponential decay produces a straight line.} 
\label{Fig2}
\end{figure}

To study the effect of magnetic fields and excess carrier density on the recombination dynamics, we measured the magnetic-field and laser-fluence dependent photoinduced transmission for Nb$_{0.5}$Ti$_{0.5}$N and NbN thin films.  The samples were fully immersed in superfluid $^4$He ($T\le2$~K) to minimize heating. At this low temperature, the thermal quasiparticle population is small (compared to the number of broken pairs at high fluence) but not zero. The field was applied parallel to the film surface to avoid the complexity of vortex effects. (See Ref.~\cite{Xi2010}). Typical results are shown in FIG.~\ref{Fig2}, where the photoinduced signal $S(t)$ (excess quasiparticle density) is plotted against delay time. At both low (FIG.~\ref{Fig2}a and FIG.~\ref{Fig2}c) and high laser fluences (FIG.~\ref{Fig2}b and FIG.~\ref{Fig2}d), a longer time is required for recombination as the magnetic field is increased. The pulse width of the synchrotron probe beam gives rise to the initial upturn in the data, which is skipped in the following data analysis.

\begin{figure}[t]
\includegraphics[width=0.3\textwidth]{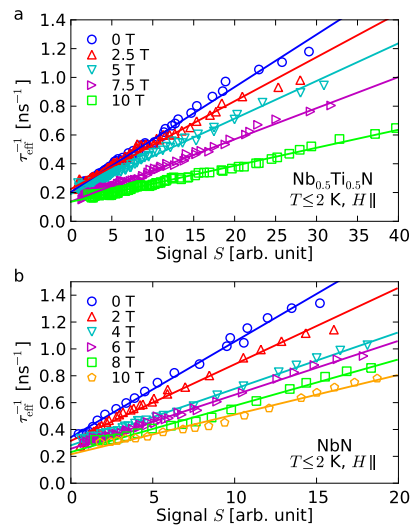}                
\caption{Effective instantaneous recombination rate vs. photoinduced transmission.  (a) For Nb$_{0.5}$Ti$_{0.5}$N, data at each field include fluences ranging from 0.4 to 10.7~nJ/cm$^2$. (b) For NbN, data at each field include fluences ranging from 2.4 to 18.1~nJ/cm$^2$ except for 8~T and 10~T, where data were collected at 18.1~nJ/cm$^2$. A 4-point moving average was performed on the data to reduce noise. The lines are linear fits to the data.} 
\label{Fig3}
\end{figure}

We have discovered a revealing perspective to display our results, shown in FIG.~\ref{Fig3}. We define an effective instantaneous recombination rate $1/\tau_{\mathrm{eff}}(t) \equiv  -[dS(t)/dt]/S(t)$ and plot $1/\tau_{\mathrm{eff}}(t)$ vs $S(t)$ at various fields and fluences. Here short times are at the right (large $S(t)$) and long times at the left. In this presentation, data at the same field but for different pump fluences scale to the same straight line. As will be shown below, this behavior is expected for bimolecular recombination where the lifetime for a given particle is proportional to the availability of other particles with which to combine.  

\begin{figure*}[t]
\includegraphics[width=0.62\textwidth]{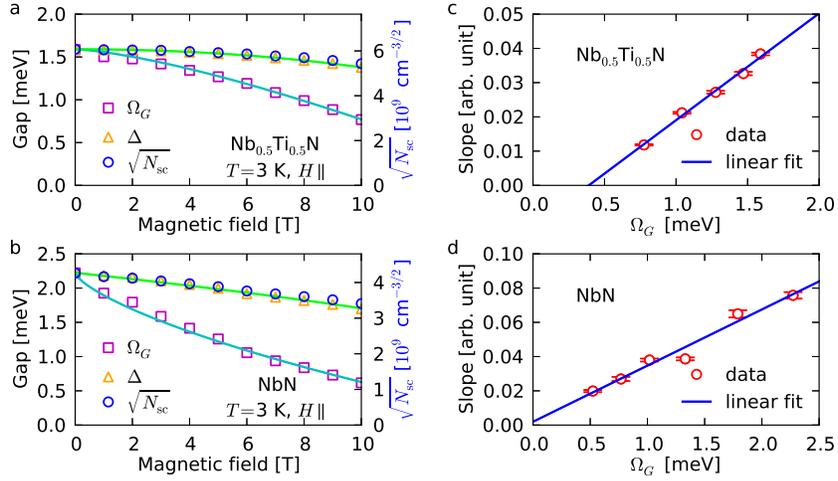}                
\caption{Panels a and b show the excitation gap, $\Omega_G$ (squares) and the pair-correlation gap $\Delta$ (triangles) for Nb$_{0.5}$Ti$_{0.5}$N and NbN, obtained from the optical conductivity (left scale). The solid lines are theoretical calculations of $\Delta$ and $\Omega_G$. The square root of the condensate density $\sqrt{N_{\mathrm{sc}}}$ (proportional to the order parameter) is shown as circles (right scale). Panels c and d show the slope extracted from FIG.~\ref{Fig3} vs. $\Omega_G$ from a and b. The error bars in both plots are calculated deviations of the slope from the linear fit in FIG.~\ref{Fig3}. The lines are linear fits to the circles. } 
\label{Fig4}
\end{figure*}

The scaling can be understood as follows. The phonon bottleneck was first discussed by Rothwarf and Taylor \cite{Rothwarf1967} using two rate equations, one for the quasiparticles and the other for the 2$\Delta$ phonons. The quasiparticles, which directly correspond to our signal $S(t)$,  follow a simple model that captures the feature of bimolecular recombination, meanwhile taking into account the phonon bottleneck. The decay rate of the total quasiparticle density $N(t)$ toward the equilibrium density is proportional to $N^2$, because recombination requires the presence of two quasiparticles. Motivated by the Rothwarf-Taylor \cite{Rothwarf1967} equations, we write
\begin{equation}
\frac{dN}{dt}=-2R(N^2-N_{\mathrm{th}}^2). \label{eq1}
\end{equation}
A thermal term $N_{\mathrm{th}}^2$ is subtracted from $N^2$, because at equilibrium $N = N_{\mathrm{th}}$ and the quasiparticle density must remain constant. The phonon bottleneck is introduced into the model through the recombination rate coefficient $R$. (See Section 1 of the Supplemental Material.) A factor of 2 is included because each recombination event depletes two quasiparticles. Now, $N(t) = N_{\mathrm{th}} + N_{\mathrm{ex}}(t)$, with  $N_{\mathrm{th}}$ the thermal density and $N_{\mathrm{ex}}(t)$ the photoinduced excess density. At a given temperature and magnetic field, $N_{\mathrm{th}}$ is time-independent, making Eq.~\eqref{eq1} become $-(dN_{\mathrm{ex}}/dt)/N_{\mathrm{ex}} = 2R(N_{\mathrm{ex}}(t) + 2N_{\mathrm{th}})$. We identify $-(dN_{\mathrm{ex}}/dt)/N_{\mathrm{ex}}$ as the effective instantaneous relaxation rate $1/\tau_{\mathrm{eff}}(t)$ defined earlier, because the photoinduced transmission $S(t)$ is proportional to the excess quasiparticle density \cite{Lobo2005}, $S = CN_{\mathrm{ex}}$, where $C$ is just a constant to convert from signal to quasiparticle density. Hence,
\begin{equation}
-\frac{1}{S(t)}\frac{dS(t)}{dt} = \frac{2R}{C}(S(t)+2S_{\mathrm{th}}), \label{eq2}
\end{equation}
with $S_{\mathrm{th}} = CN_{\mathrm{th}}$. Eq.~\eqref{eq2} is consistent with the linear behavior demonstrated in FIG.~\ref{Fig3}. The field dependence requires the prefactor $R$ to decrease with field. 

To interpret the field dependence shown in FIG.~\ref{Fig3}, it is a prerequisite to understand how the field changes the electronic states of the superconductor. If spin-orbit scattering is small, the magnetic field could make the majority of quasiparticles have one spin direction. (This is the same polarization that gives Pauli paramagnetism to metals.) Spin polarization will slow the recombination because only quasiparticles with opposite spins can recombine. A recombination model including this spin polarization effect is discussed in Section 2 of the Supplemental Material. In this case, the recombination equation remains in the same form as Eq.~\eqref{eq2}, but with the coefficient $2R/C$ replaced by $(8R/C)(N^{\uparrow}N^{\downarrow}/N^2)$, where $N^{\uparrow}$ and $N^{\downarrow}$ are respectively the densities of spin-up and spin-down quasiparticles. The quasiparticle spin-polarization factor $N^{\uparrow}N^{\downarrow}$ would depend on the magnetic field in the limit of weak spin-orbit coupling, just as in the Pauli paramagnetism of metals. According to the BCS theory, electrons form spin-singlet pairs condensed in the ground state; the spin susceptibility vanishes as the temperature approaches 0. The studies of superconductor spin susceptibility were done on thin films with thickness so small that the effect of a magnetic field on the electron orbit could be neglected. Paramagnetic splitting of the quasiparticle density of states was observed in 5 nm aluminum films in a parallel magnetic field \cite{Meservey1970}. In a study of magnetic field effects on far-infrared absorption of thin superconducting aluminum films, van Bentum and Wyder \cite{Bentum1986} concluded that paramagnetic splitting was important in their thinnest films, but did not allow for quasiparticle spin polarization. If a high degree of spin polarization existed, the recombination rate would be slowed much more than observed. However, spin-orbit scattering must be considered. Tedrow and Meservey observed the spin-state mixing in thin aluminum films due to spin-orbit scattering \cite{Tedrow1971}. They defined a spin-orbit scattering parameter $b\equiv\hbar/3\Delta\tau_{\mathrm{so}}$ to describe the degree of spin-orbit scattering, where $\tau_{\mathrm{so}}$ is the spin-orbit scattering time. They calculated that, as $b$ is increased to 0.5, the spin-up and spin-down quasiparticle density of states completely mix, leaving no clear signature of the two-peak feature in the density of states due to Zeeman splitting. Considering the short spin-orbit scattering time measured \cite{Hake1967} in NbTi, $\tau_{\mathrm{so}} = 3.0\times 10^{-14}$~s, and using the $\Delta$ of Nb$_{0.5}$Ti$_{0.5}$N and NbN, we estimate that $b = 4.2$ and 3.3 for Nb$_{0.5}$Ti$_{0.5}$N and NbN, respectively. We believe that spin is not a good quantum number in our samples, requiring us to look beyond spin polarization to understand the recombination.

In a study \cite{Xi2010} of the optical conductivity of Nb$_{0.5}$Ti$_{0.5}$N, we found that a parallel magnetic field breaks the time-reversal symmetry of the Cooper pairs and decreases the superconducting energy gap. The physics is similar to magnetic-impurity-induced pair-breaking effects, as originally formulated by Abrikosov and Gor’kov \cite{Abrikosov1961}. In a magnetic field, one must distinguish between the spectroscopic energy gap 2$\Omega_G$ and the pair-correlation gap $\Delta$. These gaps \cite{Skalski1964} are plotted in FIG.~\ref{Fig4}a as squares and triangles respectively. The real part of the optical conductivity, corresponding to the electromagnetic absorption, shows a clear suppression of the energy gap 2$\Omega_G$ with field (squares in FIG.~\ref{Fig4}a). The imaginary conductivity is a measure of the superconducting condensate density $N_{\mathrm{sc}}$, which goes as $\Delta^2$. The field dependences of $\Delta$ and of $\sqrt{N_{\mathrm{sc}}}$ (shown as circles) agree well, providing clear evidence for a weakening of superconductivity by the magnetic field. The quantities $\Omega_G$, $\Delta$, and $N_{\mathrm{sc}}$ for NbN, obtained using the same technique (in Section 3 of the Supplemental Material), are plotted in FIG.~\ref{Fig4}b. The NbN field dependence is qualitatively different from that of Nb$_{0.5}$Ti$_{0.5}$N because in this thicker film the applied field induces a spatial variation in the order parameter, making the weakening of superconductivity be proportional to the field, rather than being quadratic in field as in the much thinner Nb$_{0.5}$Ti$_{0.5}$N \cite{Parks1969}. The energy gaps will be used in the following analysis.

The field dependence, shown in FIG.~\ref{Fig3}, is dominated by the recombination rate coefficient $R$. On the one hand, by explicitly solving Eq.~\eqref{eq2} one can identify a low-fluence recombination rate $1/\tau_{\mathrm{eff}} = 4RN_{\mathrm{th}}$. (See Section 4 of the Supplemental Material.) The field dependence of the thermal quasiparticle density $N_{\mathrm{th}}$ results from the field-dependent energy gap and the quasiparticle density of states \cite{Bayrle1989}. On the other hand, the effective lifetime of the excess quasiparticles is modified from the intrinsic value $\tau_R$, and is tied to the rates at which the phonons, produced in recombination events, re-break pairs (1/$\tau_B$) or escape from the film (1/$\tau_{\gamma}$) \cite{Gray1971}. The quasi-equilibrium values of $\tau_R$ and $\tau_B$ were derived by Kaplan \textit{et al.} \cite{Kaplan1976}.  Magnetic-field-induced pair breaking decreases the spectroscopic energy gap (FIG.~\ref{Fig4}) and modifies the quasiparticle density of states, resulting in a decrease in $\tau_R$ and an increase in $\tau_B$. The field independent phonon escape time is determined by the film thickness and the acoustic mismatch between the film and the environment \cite{Chang1978}. The recombination rate coefficient $R$ (and, hence, the slope of $1/\tau_{\mathrm{eff}}$ in FIG.~\ref{Fig3}) is therefore field-dependent through $N_{\mathrm{th}}$, $\tau_R$ and $\tau_B$. (See FIG. S7 in the Supplemental Material.) The equation is involved but, when we compute the slope vs $\Omega_G$ for Nb$_{0.5}$Ti$_{0.5}$N and NbN, shown in FIG.~\ref{Fig4}c and FIG.~\ref{Fig4}d, we obtain a basically linear relation. This calculation implies a connection between the field-dependent quasiparticle recombination and the field-induced pair breaking. The linear relation can be explained by considering only the field-induced gap reduction. (See FIG. S7 in the Supplemental Material.) The finite y-intercepts in FIG.~\ref{Fig4}c and FIG.~\ref{Fig4}d are intriguing, bringing out the question of how the photoexcited quasiparticles relax in a gapless superconductor, motivating challenging experiments to probe the gapless regime.

In conclusion, our time-resolved pump-probe measurements on metallic \textit{s}-wave superconductors reveal a slowing of quasiparticle recombination in an external magnetic field. The field was aligned parallel to the thin-film sample surface, to minimize effects due to vortices. There are two possible causes of the observed slowing: field-induced spin imbalance and field-induced gap reduction. The spin imbalance is unlikely to be important in Nb$_{0.5}$Ti$_{0.5}$N and NbN due to strong spin-orbit scattering. This scenario can be tested by investigating materials with small spin-orbit scattering. The field-induced gap reduction alone can explain
quantitatively the slowing of recombination, and we conclude it to be the dominant effect observed in our experiment.
\vspace{6 mm}

We thank P. Bosland and E. Jacques for providing the samples, J. J. Tu for access to the magnet, R. P. S. M. Lobo for data acquisition software development, G. Nintzel and R. Smith for technical assistance, and P. J. Hirschfeld for discussions. This work was supported by the U.S. Department of Energy through contracts DE-ACO2-98CH10886 (Brookhaven National Laboratory) and DEFG02-02ER45984 (University of Florida), and by the National Research Foundation of Korea through Grant No. 20100008552.

\clearpage

%%%%%%%%%%%%%%%%%%%%%%%%%%%%%%%%%%%%%
%%%     Supplemental Material     %%%
%%%%%%%%%%%%%%%%%%%%%%%%%%%%%%%%%%%%%
\setcounter{figure}{0}
\setcounter{page}{1}
\pagenumbering{arabic}
\noindent
\textbf{Supplemental Material}

%%%%%%%%%%%%%%%%%%%%%%%%%
%%%     Section 1     %%%
%%%%%%%%%%%%%%%%%%%%%%%%%
\section{1. P\lowercase{honon bottleneck effect}}
In this section we show that the phonon bottleneck effect can be included in the effective recombination rate $R$ as a proportionality coefficient. Consider the Rothwarf-Taylor equations \cite{Rothwarf1967} for the coupled populations of quasiparticles and phonons in the absence of external quasiparticle injection,
\begin{equation}
\frac{dN}{dt} = -\tau_R^{-1}N^2+\tau_B^{-1}N_{\omega}, \label{eqS1}\tag{S1}
\end{equation}
\begin{equation}
\frac{dN_{\omega}}{dt} = \frac{1}{2}\tau_R^{-1}N^2-\frac{1}{2}\tau_B^{-1}N_{\omega}-\tau_{\gamma}^{-1}(N_{\omega}-N_{\omega,\mathrm{th}}),\label{eqS2}\tag{S2}
\end{equation}
where $N(t)$ and $N_{\omega}(t)$ are respectively the densities of quasiparticles and high-energy (defined as $\hbar\omega\ge 2\Delta$) phonons, $\tau_R^{-1}$ is the intrinsic quasiparticle recombination rate, $\tau_B^{-1}$ is the phonon pair-breaking rate, and $\tau_{\gamma}^{-1}$ is the phonon escape rate. (The phonons may enter the substrate, enter the helium bath, or decay anharmonically to energies $\hbar\omega<2\Delta$.) In thermal equilibrium, the quasiparticles $N(t)$ and high-energy phonons $N_{\omega}(t)$ reach time-independent equilibrium values, linked by $N_{\mathrm{th}}=\sqrt{N_{\omega,\mathrm{th}}\tau_B^{-1}/\tau_R^{-1}}$, where $N_{\mathrm{th}}$ and $N_{\omega,\mathrm{th}}$ are for the quasiparticles and high-energy phonons, respectively. The coupled non-linear equations~\eqref{eqS1} and~\eqref{eqS2} can be solved numerically, e.g., using the Runge-Kutta integration method. To illustrate the solutions, we set $\tau_R^{-1}=1$~ns$^{-1}$ and $\tau_B^{-1}=10$~ns$^{-1}$, typical values for these quantities at low temperatures \cite{Lobo2005}. Without loss of generality we set $N_{\omega,\mathrm{th}}=1$, which determines $N_{\mathrm{th}}=\sqrt{10}$. The initial value of $N$ is determined by the pump laser fluence. We consider three cases, $N(0) = 2N_{\mathrm{th}}$ for low fluence, $N(0) = 5N_{\mathrm{th}}$ for intermediate fluence, and $N(0) = 10N_{\mathrm{th}}$ for high fluence. $N_{\omega}(0)$ is set to $N_{\omega,\mathrm{th}}$. For each case, we consider a range of phonon escape rates ($\tau_{\gamma}^{-1}$) from 1~ns$^{-1}$ to 10~ns$^{-1}$, spanning the strong to the weak phonon-bottleneck regime.

\begin{figure*}[t]
\renewcommand{\thefigure}{S\arabic{figure}}
\includegraphics[width=1.0\textwidth]{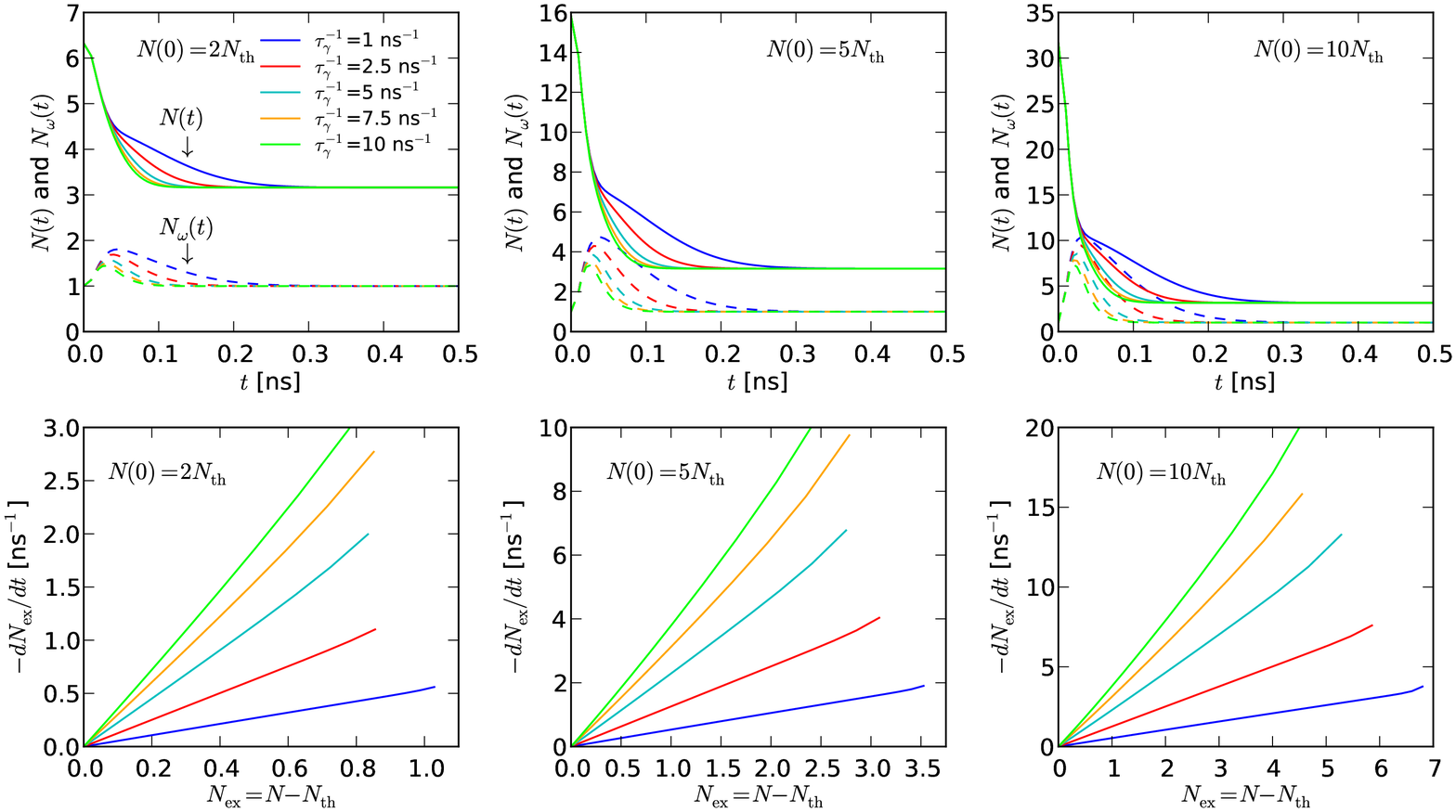}                
\caption{Numerical solutions of the Rothwarf-Taylor equations. The first row shows the solutions for the quasiparticle density $N$ and the high-energy phonon density $N_{\omega}$ at different fluences and for a range of phonon escape rate. The second row shows the rate of change of the excess quasiparticle density $-(dN_{\mathrm{ex}}/dt)/N_{\mathrm{ex}}$ as a function of $N_{\mathrm{ex}}$.} 
\label{FigS1}
\end{figure*}

The numerical solutions of the Rothwarf-Taylor equations are shown in the first row of FIG.~\ref{FigS1}. After a short period of approximately 0.05~ns, the phonon bottleneck effect becomes clear. As expected, when the phonon escape rate increases, the phonon bottleneck effect becomes weaker and recombination becomes faster. In the second row of FIG.~\ref{FigS1}, the rate of change of the excess quasiparticle density $N_{\mathrm{ex}}(t)=N(t)-N_{\mathrm{th}}$ is plotted vs. $N_{\mathrm{ex}}$. The early stage ($t<0.05$~ns) shown in the first row of FIG.~\ref{FigS1} has been skipped, because the quasiparticle and phonon populations are not yet equilibrated and it does not give information about the phonon bottleneck. Moreover, this stage is not temporally resolved in our experiment.

\begin{figure}[h!]
\renewcommand{\thefigure}{S\arabic{figure}}
\includegraphics[width=0.4\textwidth]{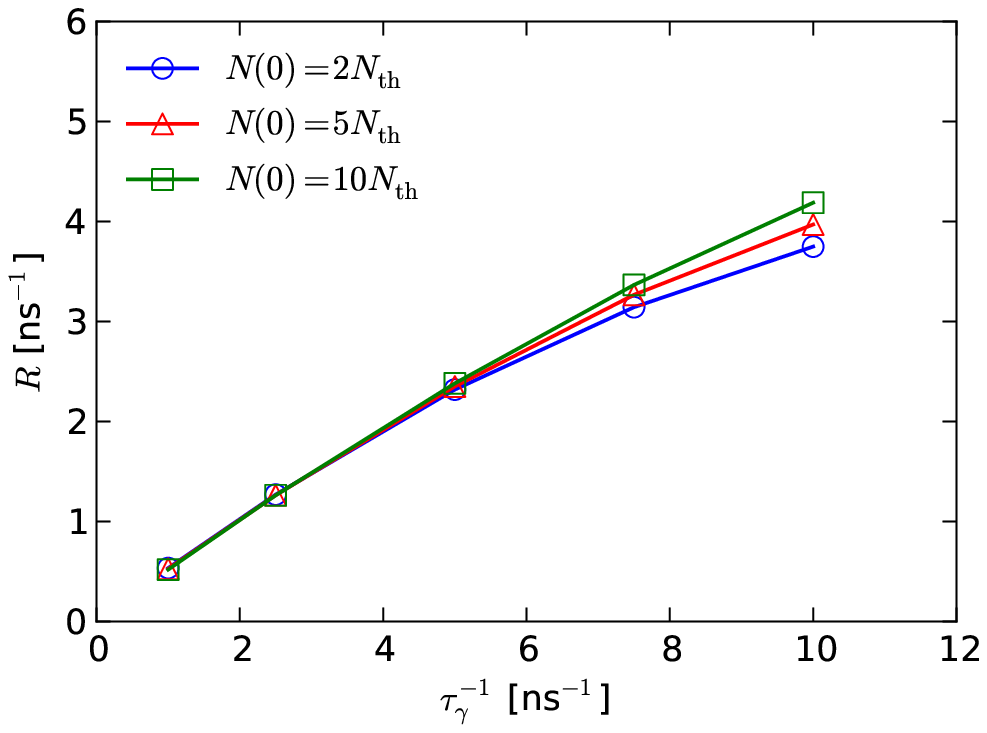}                
\caption{The effective recombination rate as a function of the phonon escape rate at low, intermediate and high fluence.} 
\label{FigS2}
\end{figure}

The quasi-linear relation shown in the second row of FIG.~\ref{FigS1} suggests we can define an effective recombination rate $R$ as $-(dN_{\mathrm{ex}}/dt)/N_{\mathrm{ex}}$. This rate, defined as the slope of $-dN_{\mathrm{ex}}/dt$ vs. $N_{\mathrm{ex}}$, is plotted vs. $\tau_{\gamma}^{-1}$ in FIG.~\ref{FigS2}. As $N(0)$ increases from 2$N_{\mathrm{th}}$ to 10$N_{\mathrm{th}}$, the relation between $R$ and $\tau_{\gamma}^{-1}$ becomes almost linear. We estimated that, for our experimental conditions, $N(0)$ in our samples was three orders of magnitude greater than $N_{\mathrm{th}}$ at the lowest laser fluence. Therefore it is safe to conclude that the effective recombination rate $R$ scales linearly with the phonon escape rate $\tau_{\gamma}^{-1}$.

%%%%%%%%%%%%%%%%%%%%%%%%%
%%%     Section 2     %%%
%%%%%%%%%%%%%%%%%%%%%%%%%
\section{2. P\lowercase{auli paramagnetism}}
Our recombination model can be extended to include the magnetic-field dependence due to quasiparticle spin polarization. In a quasiparticle recombination event, both a spin-up and a spin-down quasiparticle are needed to form a Cooper pair. We use an equation similar to the band-to-band recombination equation in a semiconductor to describe the quasiparticle recombination, 
\begin{equation}
\frac{dN}{dt}=\frac{dN^{\uparrow}}{dt}+\frac{dN^{\downarrow}}{dt} = -8R(N^{\uparrow}N^{\downarrow}-N_{\mathrm{th}}^{\uparrow}N_{\mathrm{th}}^{\downarrow}),\label{eqS3}\tag{S3}
\end{equation}
where $\uparrow$ and $\downarrow$ denote the spin-up and spin-down populations, respectively. This reduces to Eq.~(1) in the main text when the spin-up and spin-down populations are equal. At a given condition, Eq.~\eqref{eqS3} can be rewritten as $-(dN_{\mathrm{ex}}/dt)/N_{\mathrm{ex}}=8R(N^{\uparrow}N^{\downarrow})(N_{\mathrm{ex}}+2N_{\mathrm{th}})/N^2$, where $N_{\mathrm{ex}} = N-N_{\mathrm{th}}$. Relating the measured photoinduced transmission $S$ to the excess quasiparticle density, $S = CN_{\mathrm{ex}}$, where $C$ is a constant, the recombination equation can be further reduced to 
\begin{equation}
-\frac{1}{S}\frac{dS}{dt}=8\frac{R}{CN^2}(N^{\uparrow}N^{\downarrow})(S+2S_{\mathrm{th}}),\label{eqS4}\tag{S4}
\end{equation}
where $S_{\mathrm{th}}=CN_{\mathrm{th}}$.

\begin{figure}[h!]
\renewcommand{\thefigure}{S\arabic{figure}}
\includegraphics[width=0.4\textwidth]{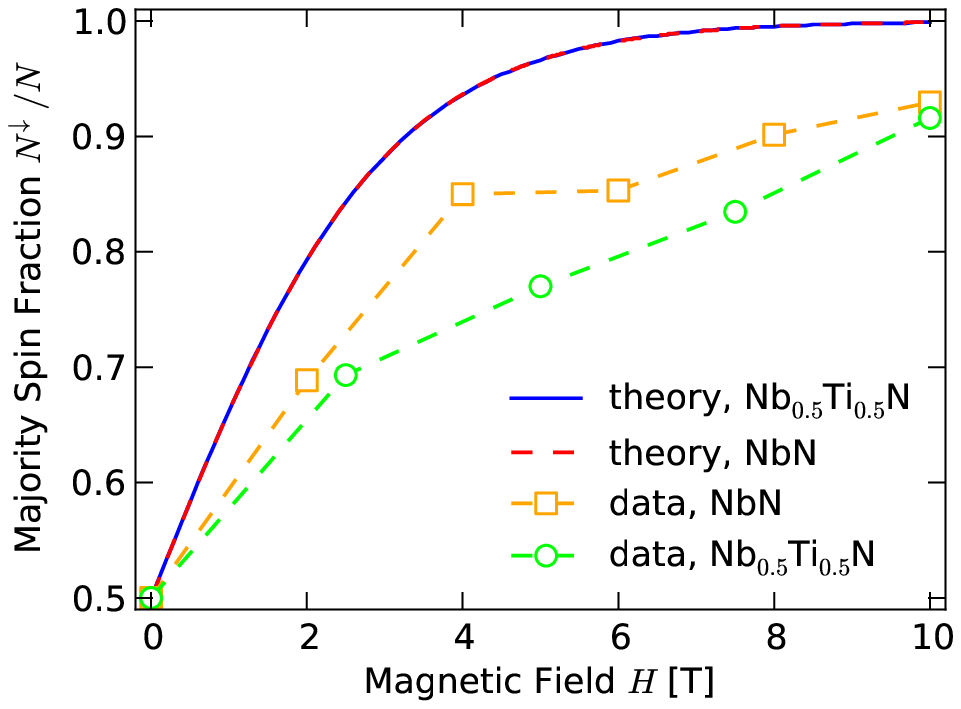}                
\caption{Majority spin fraction. The theory is based on Eqs.~\eqref{eqS5} and~\eqref{eqS6}. The data points are extracted from the slopes in FIG.~3 in the main text, all scaled so that the majority spin fraction is 1/2 at 0~T.} 
\label{FigS3}
\end{figure}

The majority spin fraction $N^{\downarrow}/N$ can be calculated from the paramagnetic model, assuming Fermi-Dirac distribution of quasiparticles $f(E)$ and the quasiparticle density of states $D(E)$ from the BCS theory,
\begin{align}
N^{\downarrow} &= 2\int_0^{\infty}f(E)D(E+\mu H)dE,\label{eqS5}\tag{S5}\\
N^{\uparrow} &= 2\int_0^{\infty}f(E)D(E-\mu H)dE.\label{eqS6}\tag{S6}
\end{align}
If we assume that the field dependence we see in FIG.~3 in the main text is only through the product $N^{\uparrow}N^{\downarrow}$, we can extract the majority spin fraction at different fields. The results are compared with the calculation in FIG.~\ref{FigS3}. Pure Pauli paramagnetism predicts a stronger magnetic field dependence than observed in our data. 

The theory, however, must consider the strong spin-orbit scattering in NbN and Nb$_{0.5}$Ti$_{0.5}$N. In the main text, we estimate the strong spin-state mixing in the Nb$_{0.5}$Ti$_{0.5}$N and NbN samples. Based on that argument and the analysis shown in this section, we expect that the spin polarization factor $N^{\uparrow}N^{\downarrow}$ is weakly dependent on the field for our samples.

%%%%%%%%%%%%%%%%%%%%%%%%%
%%%     Section 3     %%%
%%%%%%%%%%%%%%%%%%%%%%%%%
\section{3. A\lowercase{nalysis of optical conductivity for} N\lowercase{b}N}
\begin{figure}[t]
\renewcommand{\thefigure}{S\arabic{figure}}
\includegraphics[width=0.49\textwidth]{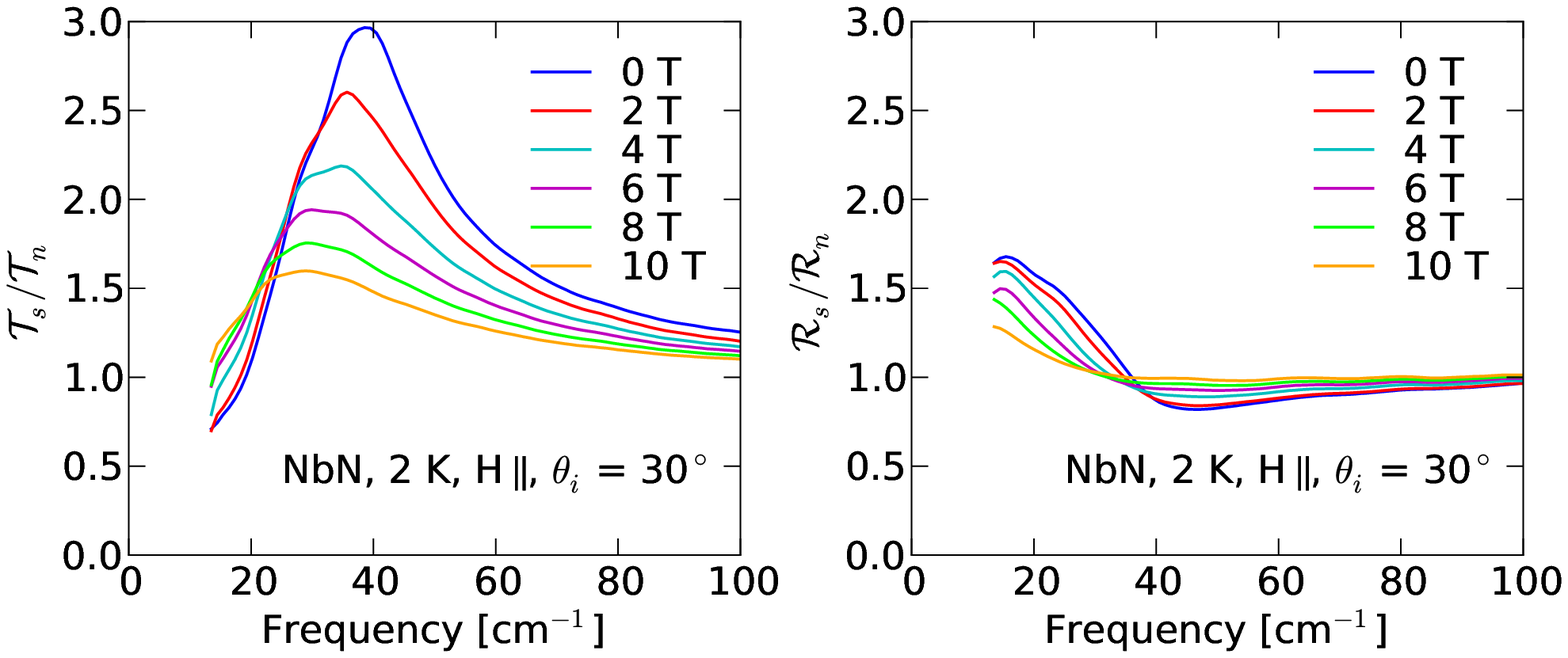}                
\caption{The transmission and reflection of NbN in parallel fields and 2~K, normalized to the corresponding normal-state values. The angle of incidence $\theta_i$ for both transmission and reflection was 30$^{\circ}$.} 
\label{FigS4}
\end{figure}
We studied the magnetic-field-induced effects in the Nb$_{0.5}$Ti$_{0.5}$N and NbN thin films using Fourier transform far-infrared spectroscopy. The experimental technique and the analysis for Nb$_{0.5}$Ti$_{0.5}$N can be found in Ref.~\cite{Xi2010}. Here we analyze the superconducting-state to normal-state transmission ratio $\mathcal{T}_s/\mathcal{T}_n$ and reflection ratio $\mathcal{R}_s/\mathcal{R}_n$ for NbN shown in FIG.~\ref{FigS4}, measured at 2~K with the magnetic field parallel to the film. The data were taken with 4~cm$^{-1}$ (0.5~meV) resolution, so that the fringes due to the multiple internal reflections in the substrate were not resolved. The angle of incidence for both transmission and reflection was 30$^{\circ}$. The NbN thin film has a normal-state conductivity $\sigma_n = 2.0\times 10^3~\mathrm{Ohm}^{-1}\mathrm{cm}^{-1}$, determined from its normal-state transmittance and thickness. The MgO substrate has a refractive index $n \approx 3.0$ and negligible absorption in the far-infrared. The zero-field gap $\Delta_0 = 17.9$~cm$^{-1}$ is obtained from fitting the zero-field optical conductivity with the Mattis-Bardeen theory. 

\begin{figure}[b!]
\renewcommand{\thefigure}{S\arabic{figure}}
\includegraphics[width=0.49\textwidth]{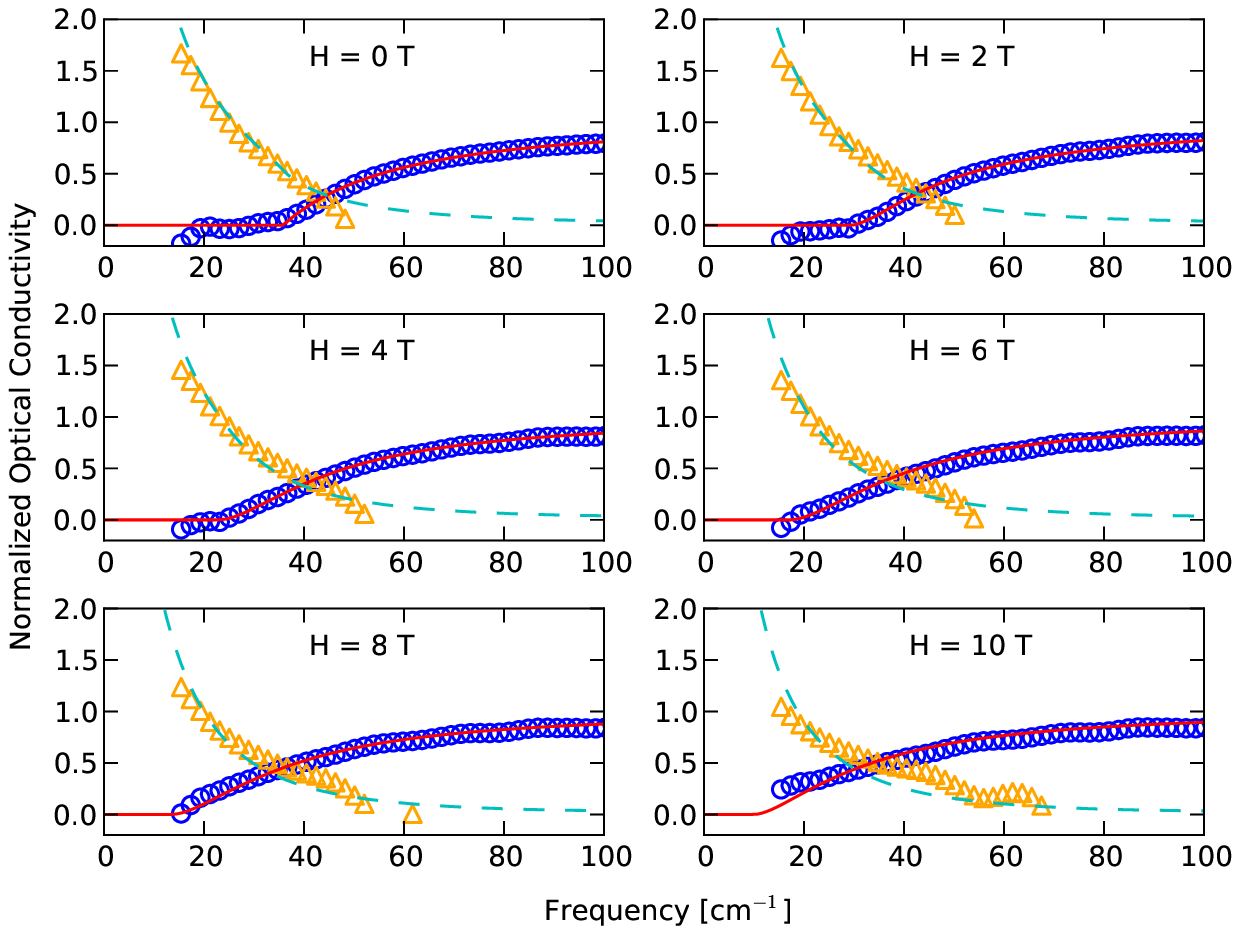}                
\caption{The real (circles) and imaginary (triangles) parts of the optical conductivity of NbN at various parallel fields and $T=2$~K, normalized to the normal-state conductivity $\sigma_n$. The solid lines are fits to $\sigma_1/\sigma_n$ using the pair-breaking theory \cite{Skalski1964}. The dashed lines show the corresponding $\sigma_2/\sigma_n$ as determined by a Kramers-Kronig transform of the fit to the real part.} 
\label{FigS5}
\end{figure}

\begin{figure}[t!]
\renewcommand{\thefigure}{S\arabic{figure}}
\includegraphics[width=0.4\textwidth]{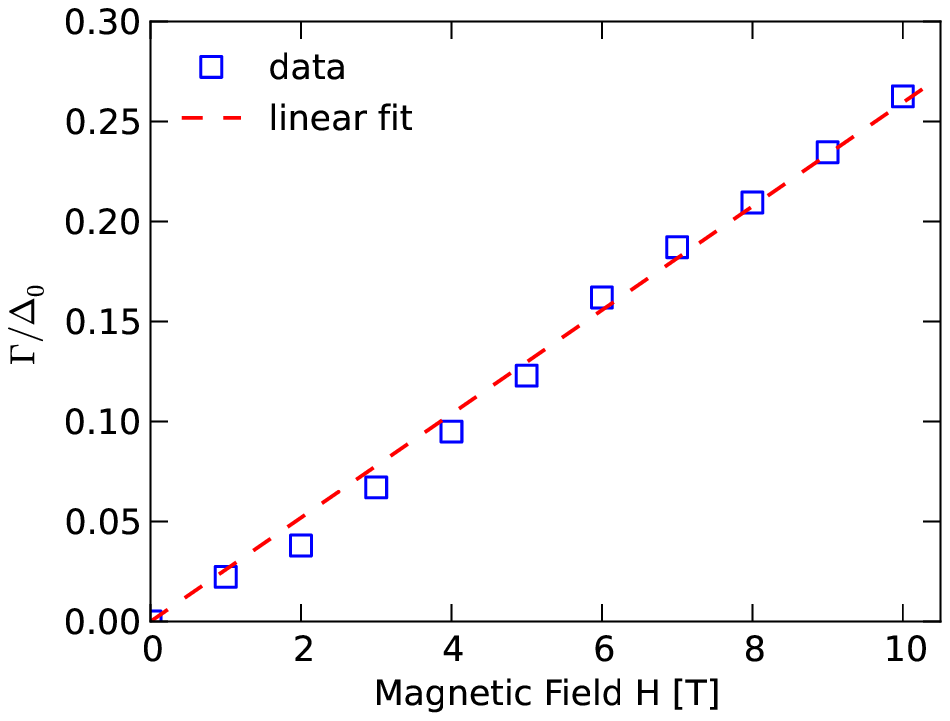}                
\caption{Field dependence of the pair-breaking parameter $\Gamma$ determined from the optical conductivity of NbN, normalized to the zero-field gap. The dashed line is a linear fit.} 
\label{FigS6}
\end{figure}
 
From the transmission and reflection ratios, we extracted the real ($\sigma_1$) and imaginary ($\sigma_2$) parts of the optical conductivity using the method discussed in Ref.~\cite{Xi2010}. The results are shown in FIG.~\ref{FigS5}. We found that the field dependence can be explained well by the pair-breaking theory. (See details of the theory in Ref.~\cite{Skalski1964}.)

The pair-breaking parameter $\Gamma$ is the only fitting parameter, describing the strength of pair breaking due to the magnetic field. Its value at different fields is shown in FIG.~\ref{FigS6}. From $\Gamma$ we calculated the pair-correlation gap $\Delta$ and the effective spectroscopic gap $\Omega_G$ using $\mathrm{ln}(\Delta/\Delta_0) = -\pi\Gamma/4\Delta$ and $\Omega_G = \Delta[1-(\Gamma/\Delta)^{2/3}]^{3/2}$. These quantities are shown in FIG.~4b in the main text. The superconducting condensate density $N_{\mathrm{sc}}$ is estimated from the below-gap part of $\sigma_2$ at $T\ll T_c$, which has the form,
\begin{equation}
\sigma_{2}(\omega) = \frac{N_{\mathrm{sc}}e^2}{m\omega},\label{eqS7}\tag{S7}
\end{equation}
where $e$ and $m$ are the electron charge and mass, respectively.

%%%%%%%%%%%%%%%%%%%%%%%%%
%%%     Section 4     %%%
%%%%%%%%%%%%%%%%%%%%%%%%%
\section{4. E\lowercase{xact solution of the recombination equation}}
The recombination equation, Eq.~(2) in the main text, links the measured $S(t)$ to the model,
\begin{equation}
-\frac{1}{S(t)}\frac{dS(t)}{dt} = \frac{2R}{C}(S(t)+2S_{\mathrm{th}}), \nonumber
\end{equation}
where $S_{\mathrm{th}} = CN_{\mathrm{th}}$. This equation has the following exact solution:
\begin{equation}
S(t) = S(0)\frac{2e^{-t/\tau}}{2+\frac{\displaystyle S(0)}{\displaystyle S_{\mathrm{th}}}(1-e^{-t/\tau})}.\label{eqS8}\tag{S8}
\end{equation}
Here $\tau$ is the effective lifetime: $\tau =1/4RN_{\mathrm{th}}$. In the low-fluence regime, especially when $S(0)\ll S_{\mathrm{th}}$, the solution is close to an exponential decay. As the fluence increases, the deviation from a simple exponential decay becomes significant.

%%%%%%%%%%%%%%%%%%%%%%%%%
%%%     Section 5     %%%
%%%%%%%%%%%%%%%%%%%%%%%%%
\section{5. E\lowercase{ffective recombination rate}}
On the one hand, the quasiparticles, interacting with phonons in the system, decay with an effective lifetime $\tau_{\mathrm{eff}}= \tau_{\gamma}+(1/2)\tau_R(1+\tau_{\gamma}/\tau_B)$, where $\tau_{\gamma}$, $\tau_R$, and $\tau_B$ are the same quantities as defined above and in the main text \cite{Gray1971}. At low temperatures, $\tau_R\gg\tau_{\gamma}$ and $\tau_{\gamma}\gg\tau_B$ \cite{Lobo2005}. The effective lifetime can be approximated as $\tau_{\mathrm{eff}}=\tau_R\tau_{\gamma}/2\tau_B$. On the other hand, by solving the recombination equation proposed in the main text, one can identify an effective quasiparticle lifetime $\tau_{\mathrm{eff}}=1/4RN_{\mathrm{th}}$, as shown in the previous section. As a result, 
\begin{equation}
R=\frac{\tau_B}{2\tau_{\gamma}\tau_RN_{\mathrm{th}}}=\frac{\tau_{\gamma}^{-1}\tau_R^{-1}}{2\tau_B^{-1}N_{\mathrm{th}}}. \label{eqS9}\tag{S9}
\end{equation}
The phonon escape rate $\tau_{\gamma}^{-1}$ is expected to be independent of the gap, as discussed in the main text. A theory for $\tau_R^{-1}$ and $\tau_B^{-1}$ has been given by Kaplan \textit{et al}. \cite{Kaplan1976},
\begin{align}
\tau_R^{-1}&(\omega)=\frac{\tau_0^{-1}}{(k_BT_c)^3[1-f(\omega)]}\int_{\omega+\Delta}^{\infty}\frac{\Omega^2(\Omega-\omega)}{\sqrt{(\Omega-\omega)^2-\Delta^2}}\nonumber\\
&\cdot\left[1+\frac{\Delta^2}{\omega(\Omega-\omega)}\right][n(\Omega)+1]f(\Omega-\omega)d\Omega,\label{eqS10}\tag{S10}
\end{align}
\begin{align}
\tau_B^{-1}&(\Omega) = \frac{\tau_{0,\mathrm{ph}}^{-1}}{\pi\Delta(0)}\int_{\Delta}^{\Omega-\Delta}\frac{\omega}{\sqrt{\omega^2-\Delta^2}}\frac{\Omega-\omega}{\sqrt{(\Omega-\omega)^2-\Delta^2}}\nonumber\\
&\cdot\left[1+\frac{\Delta^2}{\omega(\Omega-\omega)}\right][1-f(\omega)-f(\Omega-\omega)]d\omega,\label{eqS11}\tag{S11}
\end{align}
where $\tau_0^{-1}$ and $\tau_{0,\mathrm{ph}}^{-1}$ are respectively the characteristic lifetimes of the quasiparticles and phonons, determined by the electron-phonon coupling and phonon density of states. The quantities $f$ and $n$ are the Fermi-Dirac and Bose-Einstein distribution functions, respectively. $\tau_R^{-1}$ should be evaluated at the quasiparticle gap energy $\omega=\Delta$ and $\tau_B^{-1}$ should be evaluated at the phonon energy $\Omega = 2\Delta$ for the estimation of their near-equilibrium values. For $\tau_R^{-1}$
\begin{align}
\tau_R^{-1}(\Delta)&\approx \frac{\tau_0^{-1}}{(k_BT_c)^3}\int_{2\Delta}^{\infty}d\Omega\frac{\Omega^3 e^{-(\Omega-\Delta)/k_BT}}{\sqrt{(\Omega-\Delta)^2-\Delta^2}}\nonumber\\
&= \frac{\tau_0^{-1}e^{-\Delta/k_BT}}{(k_BT_c)^3}\int_0^{\infty}dx\frac{(x+2\Delta)^{5/2}}{x^{1/2}}e^{-x/k_BT},\label{eqS12}\tag{S12}
\end{align}
in which we have replaced the Bose factor $n(\Omega)$ and the Fermi factor $f(\Delta)$ by their low-temperature values. Because only small $x$ in the integrand contributes significantly to the integral, an expansion of the numerator yields
\begin{align}
\tau_R^{-1}(\Delta)&\approx \frac{\tau_0^{-1}e^{-\Delta/k_BT}}{(k_BT_c)^3}(2\Delta)^{5/2}\Big[\int_0^{\infty}dx x^{-1/2}e^{-x/k_BT}\nonumber\\
&\qquad\qquad\qquad\qquad+\frac{5}{4\Delta}\int_0^{\infty}dx x^{1/2}e^{-x/k_BT}\Big]\nonumber\\
& =  \frac{\tau_0^{-1}e^{-\Delta/k_BT}}{(k_BT_c)^3}(2\Delta)^{5/2}\sqrt{2\pi k_BT}\left(1+\frac{5k_BT}{8\Delta}\right).\label{eqS13}\tag{S13}
\end{align}
\begin{figure}[ht!]
\renewcommand{\thefigure}{S\arabic{figure}}
\includegraphics[width=0.4\textwidth]{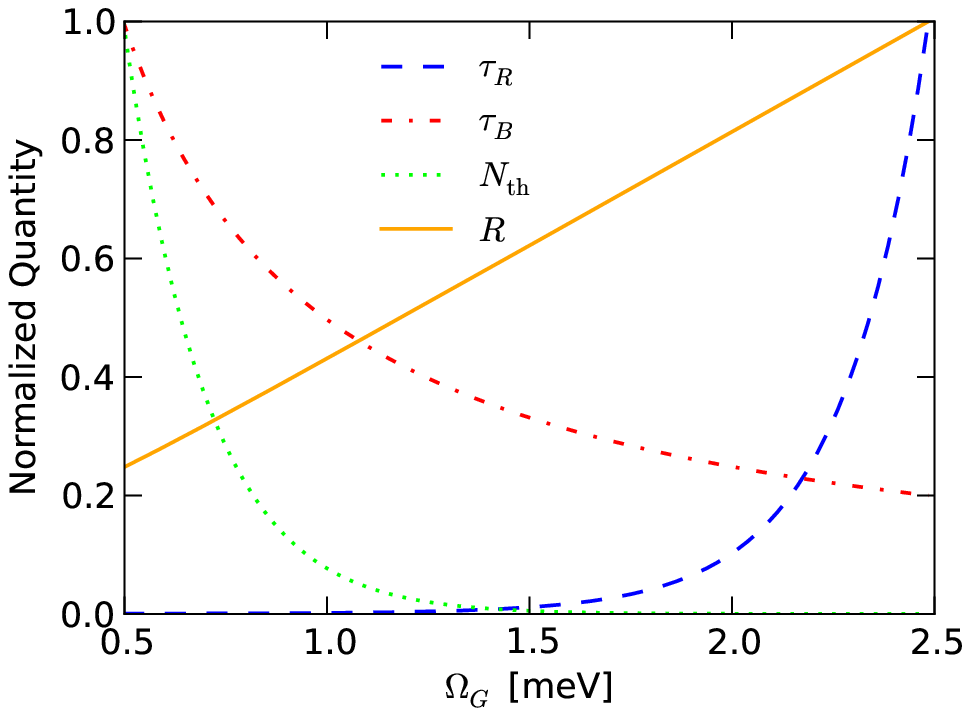}                
\caption{The dependence of $\tau_R$, $\tau_B$, $N_{\mathrm{th}}$, and $R$, on the spectroscopic gap, calculated directly from Eqs.~\eqref{eqS10},~\eqref{eqS11},~\eqref{eqS15}, and~\eqref{eqS9} without approximations.} 
\label{FigS7}
\end{figure}

\noindent
The phonon pair-breaking rate $\tau_B^{-1}$ has a simple form for its near-equilibrium state, given in Ref.~\cite{Kaplan1976} as 
\begin{equation}
\tau_B^{-1}(2\Delta)=\tau_{0,\mathrm{ph}}^{-1}\frac{\Delta}{\Delta_0}[1-2f(\Delta)]\approx \tau_{0,\mathrm{ph}}^{-1}\frac{\Delta}{\Delta_0}.\label{eqS14}\tag{S14}
\end{equation}
The quasiparticle density is 
\begin{equation}
N_{\mathrm{th}}\approx N(0)\sqrt{2\pi\Delta k_BT}e^{-\Delta/k_BT}.\label{eqS15}\tag{S15}
\end{equation}
Substituting Eqs.~\eqref{eqS13}--\eqref{eqS15} into Eq.~\eqref{eqS9} yields
\begin{equation}
R\approx\frac{2\sqrt{2}\Delta_0}{N(0)}\frac{\tau_{\gamma}^{-1}\tau_0^{-1}}{\tau_{0,\mathrm{ph}}^{-1}}\frac{1}{(k_BT_c)^3}\left(1+\frac{5k_BT}{8\Delta}\right)\Delta.\label{eqS16}\tag{S16}
\end{equation}
In the context of pair breaking, $\Delta$ should everywhere be replaced by the field-dependent spectroscopic gap $\Omega_G$.

The gap dependence of $\tau_R$, $\tau_B$, $N_{\mathrm{th}}$, and $R$ can also be numerically evaluated directly using Eqs.~\eqref{eqS10},~\eqref{eqS11},~\eqref{eqS15}, and~\eqref{eqS9} without approximations. The results are shown in FIG.~\ref{FigS7}, confirming the linear relation between $R$ and $\Omega_G$, discussed in the main text and shown in FIG.~4c and 4d.

\end{document}